\documentstyle[epsf]{elsart}
\begin{document}

\def\dirac{{\bf \rm D}\!\!\!\!/\,}
\def\wilson{{\bf \rm W}}
\def\ham{{\bf \rm H}}
\def\mbham{{\cal H}}
\def\bmat{{\bf \rm B}}
\def\cmat{{\bf \rm C}}

\def\lpmb#1{\mbox{\boldmath$#1$}}

\def\pmb#1{\setbox0=\hbox{$#1$}
\kern-.019em\copy0\kern-\wd0
\kern.03em\copy0\kern-\wd0
\kern-.025em\raise.020em\box0 }

\begin{frontmatter}

\begin{flushright}
{\normalsize FSU-SCRI-97-128}\\
{\normalsize UW-PT-97-26}\\
\end{flushright}

\title{ 
Probing the Region of Massless Quarks in Quenched Lattice QCD
using Wilson Fermions
}
\author{
Robert G. Edwards, Urs M. Heller, Rajamani Narayanan}
\address{
SCRI, The Florida State University, 
Tallahassee, FL 32306-4130, USA}
\author{Robert L. Singleton Jr.}
\address{
Department of Physics, University of Washington,
Box 351560, Seattle, WA 98195-1560, USA
}

\begin{abstract}
We study the spectrum of $\ham(m)=\gamma_5\wilson(-m)$ with
$\wilson(m)$ being the Wilson-Dirac operator on the lattice with
bare mass equal to $m$. The background gauge fields are generated
using the SU(3) Wilson action at $\beta=5.7$ on an
$8^3\times 16$ lattice. We find evidence that the spectrum of
$\ham(m)$ is gapless for $1.02 < m < 2.0$, implying that the
physical quark is massless in this whole region. 
\end{abstract}

\end{frontmatter}

{\bf PACS \#:}  11.15.Ha.\hfill\break
{Key Words:} Lattice QCD, Wilson fermions. 

\section{Introduction}

The conventional wisdom obtained from
lattice simulations of QCD using Wilson fermions
at various values of gauge couplings $\beta$ and bare quark mass $m$
is that there is a critical line 
$m_c(\beta)$
where the physical quark is massless. This critical line 
is not universal, since different definitions
for what one means by massless quark 
give different results for $m_c(\beta)$ at finite $\beta$
because the lattice theory has artifacts that only vanish in
the continuum limit.
One method is to define $m_c(\beta)$ as the line where the pion mass
is zero.
Such a definition can only be implemented by extrapolating from values
of $m$ where the pion mass is substantially away from zero.
In 
dynamical simulations, critical slowing down makes it hard to generate
enough independent configurations to obtain a good estimate of the
pion mass when it is small.
In quenched simulations, there is no problem
of critical slowing down in the generation of gauge field configurations,
but it becomes difficult to compute quark propagators below a certain
pion mass due to the presence of small eigenvalues in the spectrum of
the Wilson-Dirac operator. Thus, as already mentioned,
one relies on an extrapolation to
estimate $m_c(\beta)$ using data for
pion masses greater than $0.5m_\rho$~\cite{pion}.

The chiral symmetry of
QCD with massless quarks is expected to be spontaneously broken.
In the continuum, the spectral density $\rho(\lambda)$
of $\gamma_5\dirac$ 
is symmetric, and spontaneous breakdown of chiral symmetry is due to
a non-zero value for $\rho(0)$~\cite{bc}.
Furthermore, gauge fields
with nontrivial topology will result in exact zero eigenvalues of $\dirac$
with a specific chirality~\cite{index}. 
To study the spontaneous breakdown of
chiral symmetry on the lattice using Wilson fermions,
and also to compute the topological charge of the gauge fields
using Wilson fermions as a probe, we consider
$\ham(m)=\gamma_5\wilson(-m)$, where $\wilson(m)$ is the standard Wilson-Dirac
operator with bare mass equal to $m$.
The operator $\ham(m)$ can be viewed as the Hamiltonian 
describing the evolution of a four dimensional
fermion in a fifth direction. The evolution is in a gauge field
background that is independent of the fifth direction,
and this Hamiltonian plays a central role in the definition of the chiral
determinant~\cite{over}.
Level crossings in the spectrum of
$\ham(m)$ as a function of $m$ correspond to the presence of
topological objects in the background gauge field 
configuration with the net level crossing being the topological
charge of the gauge field~\cite{over}.
Level crossings at finite values of $\beta$
will not occur at a single value of $m$, but will instead occur
over a region in $m$~\cite{su2_top}. 
This is expected to be the case in both pure gauge theories
and QCD where the fermion dynamics is included. As such,
the spectrum of
$\ham(m)$ will not have a gap 
in a region \hbox{$m_1(\beta) < m < m_2(\beta)$}.
In this region, the spectral density in the thermodynamic limit is 
expected to have a finite density
of eigenvalues at zero leading to a spontaneous
breakdown of chiral symmetry~\cite{pfb}. 
It is natural to say that the physical quark mass is zero in the
whole region $m_1(\beta) < m < m_2(\beta)$.
The existence of a region as opposed to a single point
is due to finite lattice spacing, and we expect both $m_1(\beta)$ and
$m_2(\beta)$ to go to zero as $\beta$ goes to infinity. 
Therefore, one does not have a critical line $m_c(\beta)$
where the physical quark is massless. Instead, one has a region in
the $(\beta,m)$ plane bounded by the lines $m_1(\beta)$ and $m_2(\beta)$
where the physical quark is massless. The critical line
$m_c(\beta)$ obtained from the
extrapolation of
pion masses measured in the region $m < m_1(\beta)$ will
most likely lie in the
region between $m_1(\beta)$ and $m_2(\beta)$, but it need not
coincide with $m_1(\beta)$.
If $m_c(\beta)$ is the only point where the physical
quark mass is zero, then we have a region of bare quark masses
between $m_1(\beta)$ and $m_c(\beta)$ where the physical quark
mass is positive but the spectrum does not have a gap. Because
of low lying eigenvalues in the region $m_1(\beta) < m < m_c(\beta)$,
it is difficult to measure the propagators of mesons. 
Low lying eigenvalues make it numerically difficult
to compute the propagator, and even if one succeeds in computing
the inverse, violent fluctuations in the propagator over
the ensemble of gauge fields make it prohibitive
to get a good statistical estimate of the propagator. Therefore,
in the region $m_1(\beta) < m < m_c(\beta)$,
one cannot extract the pion mass by conventional means
and show that it is non-zero.
A method has been recently invented to remove the low lying
eigenvalues in the region $m_1(\beta) < m < m_c(\beta)$~\cite{fermi}
under the assumption that the low lying eigenvalues in the above
region are lattice artifacts. This can be thought of as a non-local
modification of the Wilson-Dirac operator; however, it might
enable one to compute the pion mass in the region
\hbox{$m_1(\beta) < m < m_c(\beta)$}. 
Gauge field configurations with
different physical properties are responsible for the low lying
eigenvalues at different values of $m$ and all these gauge field
configurations in the ensemble
are equally important and collectively give rise
to the existence of the chiral condensate. 
Therefore, it is not
appropriate to remove the low lying eigenvalues in the region
of $m_1(\beta) < m < m_c(\beta)$. Instead, we take the point of view
that the physical quark is massless in the whole region
$m_1(\beta) < m < m_2(\beta)$, which shrinks to the point $m=0$
as $\beta$ goes to infinity.

Our aim in this paper is to present some evidence in support
of the picture outlined in the previous paragraph. To this end, we
focus on pure gauge field configurations
generated using the SU(3) 
Wilson action at $\beta=5.7$ on an $8^3\times 16$
lattice. Using an ensemble of 50 configurations, we will show that
$\ham(m)$ is gapless in the mass range between $m_1=1.02$ and $m_2=2.0$,
indicating that the
physical quark is massless in this whole region.
We do not consider 
$m > 2$ since this is a region of $m$ 
which is clearly unphysical
in the continuum limit.~\footnote{In fact,
$m>1$ does not yield a unitary
theory in the continuum limit, but we are, however,
away from the continuum limit
here. At $m=2$ the free fermion has four poles in the propagator 
due to ``doublers", and since we are interested in the gap in the
spectrum of the $\ham(m)$ we avoid the region of $m>2$.}
Using the method proposed in Ref~\cite{pfb},
we will show that chiral symmetry is spontaneously broken 
in this whole region
of $m$.
For the quenched approximation on a $16^3\times 32$ lattice at
$\beta=5.7$, the data of Ref.~\cite{gf11} at $m<1$ yield a 
an extrapolated value of $m_c=1.047$ where the pion mass 
becomes zero.
This is inside the gapless region.
We will show that the pion propagator for a given
gauge field background can be estimated
quite well in this region by using the spectral representation for
$\ham(m)$ and restricting the spectral sum to the low lying eigenvalues.
This will enable us to compute the pion propagator more efficiently
than with conventional means since we can sum over all possible source points
on the lattice and effectively restore translational invariance. 
The resulting propagator, when averaged over an ensemble of configurations,
will still have large fluctuations due to the variation in the magnitude of
the low lying eigenvalues. 
But close to $m_1=1.02$ the fluctuations will not be large, 
 and we will see that the pion propagator in the region
from $m_1$ to $m_c$ is not described by an exponential decay but rather
by a power law, providing further evidence that the physical quark
described by the Wilson-Dirac fermion is massless in this region.
A little more physical insight on the properties
of the gauge fields responsible for the low lying
eigenvalues will be gained by looking at the eigenvectors
of $\ham(m)$ corresponding to these low lying eigenvalues. In particular,
the eigenvectors corresponding to zero eigenvalues of $\ham(m)$ have
a natural association with localized objects on the lattice, and
the eigenvectors themselves should be localized. In the sample of
50 configurations, we have nine such zero eigenvalues of $\ham(m)$
for $m < 1.047$. We will show that the corresponding eigenvectors 
are all localized.
The zero spatial momentum
component of these eigenvectors has a large extent in time, occupying
anywhere from seven to ten lattice spacings. A detailed study of
the level crossing over the whole range of $m$ from $m=1$ to $m=2$
yields information about the topological content of the gauge 
field, including the size distribution of the
topological objects on the lattice. The analysis of the size
distribution shows that larger objects cross at smaller values of
$m$, indicating that the crossings at
$m < m_c$ should indeed correspond to larger objects than the ones
associated with crossings at $m > m_c$. 

This paper is organized as follows. In the next section
we review the connection between the level crossings of $\ham(m)$
and topology. We also review the method to measure the chiral condensate
using a term that breaks parity~\cite{pfb}. This method measures
the chiral condensate arsing from spontaneous breaking only and gives 
a zero result unless the spectrum is gapless.
In section 3, we
measure the chiral condensate using the spectral representation
of $\ham(m)$ and show that there is a non-zero chiral condensate
in the whole region from $m_1=1.02$ to $m_2=2.0$. We also obtain 
the spectral density as a function of $m$ and show that the
gap closes at $m_1=1.02$, and we find that it
remains closed for all higher values of $m$
up to the largest value we studied, namely $m_2=2$. 
This is consistent with a non-vanishing chiral condensate in this
whole region of $m$. In section 4, we focus on the region from
$m_1=1.02$ to $m_c=1.047$. In this region we compare the shape of
the eigenvectors very close to the crossings points with the pion 
propagators
at the same value of $m$. We find excellent agreement, indicating
that the pion propagator is dominated by that particular eigenvector.
By employing the spectral representation of $\ham(m)$ and taking a
restricted spectral sum, we show that one can obtain the pion propagator
quite well over a region around the crossing point. We then use the
restricted spectral sum to compute the average pion propagator 
over the entire ensemble and show that the resulting pion propagator
is not described by an exponential decay. 
Our conclusions are presented in section 5.

\vfill\eject

\section{Properties of the spectral flow of $\pmb{\ham(m)}$}

We start this section by writing down the central quantity in the
paper, namely the Wilson-Dirac Hermitian
Hamiltonian $\ham(m)$ on
the lattice in a fixed gauge field background. Upon suppressing the
dependence on the SU(3) background gauge field $U$,
$\ham(m)$ in the chiral
basis is
\begin{equation}
\ham(m) = \pmatrix{\bmat - m & \cmat \cr \cmat^\dagger & -\bmat + m};
\ \ \ \ \ 0\le m \le 2,
\label{eq:hamil}
\end{equation}
where
\begin{eqnarray}
\cmat_{i\alpha,j\beta}(k,k^\prime) &=& {1\over 2}
\sum_\mu \sigma_\mu^{\alpha\beta} \bigl[
U^{ij}_\mu(k)\delta_{k^\prime,k+\hat\mu} - (U^\dagger_\mu)^{ij}(k^\prime)
\delta_{k,k^\prime+\hat\mu} 
\bigr] \\
\label{eq:cmat}
\bmat_{i\alpha,j\beta}(k,k^\prime) &=& {1\over 2}\delta_{\alpha,\beta}
\sum_\mu \bigl[ 2\delta_{ij}\delta_{kk^\prime}-
U^{ij}_\mu(k)\delta_{k^\prime,k+\hat\mu} - (U^\dagger_\mu)^{ij}(k^\prime)
\delta_{k,k^\prime+\hat\mu}\bigr]
\quad
\label{eq:bmat}
\end{eqnarray}
\begin{equation}
\sigma_1=\pmatrix{0 & 1\cr 1 & 0 \cr}; \ \ 
\sigma_2=\pmatrix{0 & -i\cr i & 0 \cr}; \ \ 
\sigma_3=\pmatrix{1 & 0\cr 0 & -1 \cr}; \ \ 
\sigma_4=\pmatrix{i & 0\cr 0 & i \cr} \ ,
\label{eq:pauli}
\end{equation}
and where $k$ and $k'$ label the lattice sites, $\alpha$ and
$\beta$ are two component spinor indices, and $i$ and $j$ 
are three component color indices. 

The fermionic action in terms of $\ham(m)$ is 
\begin{equation}
S_f=\psi^\prime \ham(m) \psi ,
\label{eq:action}
\end{equation}
with $\psi^\prime$ related to the standard $\bar\psi$ by $\bar\psi=
\psi^\prime\gamma_5$. The fields $\psi^\prime$ and $\psi$ are
independent Grassmann variables. 
The choice for the sign of $m$ differs from the standard one. In
the free theory ($U=1$), the action with $-m$ describes a free fermion
with mass equal to $m$. 

The operator $\ham(m)$ can also be viewed as the single particle
Hamiltonian describing the propagation of a fermionic field in
the fifth direction under a background gauge field that is static
in that direction. The many-body Hamiltonian describing the
fermionic evolution in the fifth direction is
\begin{equation}
\mbham(m) = a^\dagger \ham(m) a  ,
\label{eq:many_body}
\end{equation}
where $a$ and $a^\dagger$ are canonical fermion annihilation and
creation operators for Dirac fermions.
The chiral determinant on the lattice, which is the result of
integrating out a single Weyl fermion in a fixed gauge background,
is given by the overlap~\cite{over}
\begin{equation}
\det \cmat \equiv \langle 0 - | 0 + \rangle \ ,
\label{eq:overlap}
\end{equation}
where $|0\pm\rangle$ are the ground states of $\mbham(\pm m)$ for
some fixed value of $m$ in the range $(0,2)$. All choices of $m$
in this range give
rise to the same chiral theory in the continuum limit but will
have different lattice spacing effects away from the continuum.

Equation (\ref{eq:overlap}) along with (\ref{eq:many_body}) provides an
immediate connection between level crossings in the
spectral flow of $\ham(m)$ as a function of $m$ and the global
topology of the background gauge field~\cite{over}.
On a finite lattice containing $V$ sites,
$\ham(m)$ is a $2K\times 2K$ matrix with
$K=6V$. The ground states $|0\pm \rangle$ are obtained
by filling all the negative energy states of $\ham(\pm m)$.
For an arbitrary gauge field background, one can rigorously show
that $\ham(-m)$ always has $K$ negative energy states for all
values of $m>0$. Consider a background gauge field for which
$\ham(m)$ has $(K-Q)$ negative
eigenvalues with $Q>0$. 
Then $\det\cmat$ as defined in (\ref{eq:overlap}) is
zero and $\langle 0- | a_1^\dagger \cdots a_Q^\dagger | 0+ \rangle$
is non-zero, implying that $Q$ fermions of a fixed chirality
are created in this gauge background. 
Such a gauge field configuration
on the lattice will then be said to carry a topological charge of
$Q$ in analogy with the continuum index theorem~\cite{index}.
Examples of such gauge field configurations can be constructed
on the lattice~\cite{over}. Since $\ham(-m)$ has $K$ negative
energy states, while $\ham(m)$ has $(K-Q)$ negative energy states,
the spectral flow of $\ham(m)$ as a function of $m$ should have
a net level crossing of $Q$ (from below zero to above zero)
as one goes from $-m$ to $m$. 
Therefore $\ham(m)$ should have
at least $Q$ zero eigenvalues, and these will generically occur
at $Q$ different values of $m$. Typically, $n_+$ levels will
cross from below zero to above zero, and $n_-$ levels will
cross from above zero to below zero, with the topological
charge being given by $Q=n_+-n_-$.
Then $\ham(m)$ will have $(n_+ + n_-)$ zero eigenvalues
at $(n_+ + n_-)$ different values of $m$. 

It is instructive
to write down the zero eigenvalue equation of $\ham(m)$
using (\ref{eq:hamil}):
\begin{equation}
(\bmat - m) u + \cmat v = 0;\ \ \ \
\cmat^\dagger u -(\bmat -m )v = 0;\ \ \ \ u^\dagger u + v^\dagger v=1.
\label{eq:zero}
\end{equation}
Trivial manipulation of the above equation results in the
following necessary condition for a zero eigenvalue~\cite{over}
\begin{equation}
u^\dagger \bmat u + v^\dagger \bmat v = m.
\label{eq:size}
\end{equation}
Since $\bmat$, defined in (\ref{eq:bmat}), is positive for all
values of the background gauge field, it follows that
zero eigenvalues occur only for $m>0$~\cite{over}. 
As one takes a lattice gauge field and interpolates to the
continuum, all solutions to (\ref{eq:zero}) that occurred
at some value of $m>0$ will now occur at
$m=0$, since the norm of $\bmat$ on any state goes to zero
in the continuum. Level crossings
on the lattice will occur at different values of $m$
because the eigenvectors associated with
the level crossings have a typical size
inside which they are localized. Eigenvectors with larger
sizes will vary slower across the lattice than eigenvectors
with smaller sizes, implying that $m$ as given in (\ref{eq:size})
will be bigger for smaller size objects. Since we expect 
topological objects of varying sizes in an ensemble of gauge
fields generated at some finite gauge coupling on the lattice,
we expect level crossings to occur in a region of $m$.

Gauge fields generated at a finite gauge coupling on the lattice
have a finite correlation length in lattice units. One can define
the spectral distribution of $\ham(m)$ in the
thermodynamic limit, in which the number of lattice sites becomes
large, as the one obtained by
considering a large ensemble of gauge field configurations
generated on a lattice with all linear sizes greater than the
correlation length. The spectrum of $\ham(m)$ corresponding to each gauge
field configuration in the ensemble will be discrete, but the
spectral distribution for an infinite ensemble
is expected to be continuous. We denote the continuous spectral distribution
by $\rho(\lambda;m)$. This will be an even
function of $\lambda$ since for every eigenvalue
$\lambda$ of $\ham(m)$ in a fixed gauge field 
configuration there is an eigenvalue $-\lambda$ for
the Hamiltonian associated with the parity transformed
gauge field on the lattice\footnote{Lemma 4.4 in Ref~\cite{over}
provides a proof of this statement.}. If $\rho(0;m)$ is non-zero,
then $\ham(m)$ does not have a gap at that value of $m$.
Low lying eigenvalues of $\ham(m)$ and level crossings in
$\ham(m)$ can give rise to a non-zero value for $\rho(0;m)$.
One can establish a non-zero chiral condensate if $\rho(0;m)$
is non-zero by adding a term $ih\psi^\prime \psi$ to the
fermionic action in (\ref{eq:action})~\cite{pfb}.
In the presence of this term,
\begin{equation}
i\langle \psi^\prime \psi\rangle_U = {1\over 2K} \sum_n {h\over 
\lambda^2_n(m) + h^2}
\label{eq:chiralU}
\end{equation}
in a fixed gauge field background $U$, where
$2K$ is the total number of degrees of freedom.
The right hand side is obtained by taking the average
over $U$ and its parity transformed partner. To obtain the chiral
condensate, one has to compute the ensemble average in the limit
of infinite volume at finite values of $h$ and extrapolate to the
limit of vanishing $h$.
In this limit, we obtain a connection between
the chiral condensate and the non-zero value of $\rho(0;m)$~\cite{pfb},
namely,
\begin{equation}
i\langle \psi^\prime \psi \rangle = \pi\rho(0;m) \ .
\label{eq:chiral}
\end{equation}
Note that a non-zero value for $i\langle \psi^\prime \psi \rangle$
comes solely from $\rho(0;m)$. This is the advantage of measuring
the chiral condensate using $i\langle \psi^\prime \psi \rangle$.
Existence of the chiral condensate can be obtained by an
explicit computation of (\ref{eq:chiralU}) over an ensemble
of configurations as a function of $h$.

\section{Evidence for a chiral condensate over a region of $\pmb{m}$}

\begin{figure}
\epsfxsize=3.5in
\centerline{\epsffile{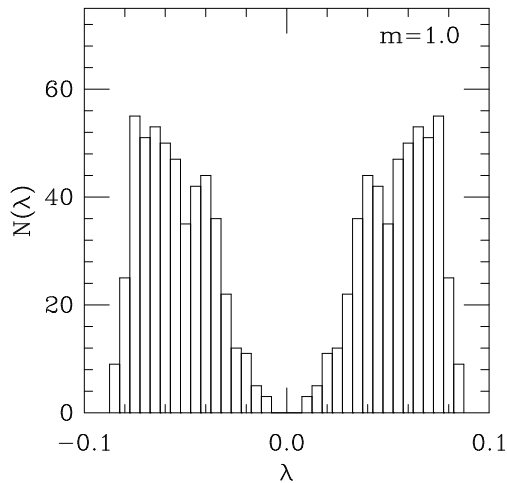}}
\caption{\tenrm
Spectral distribution of $\ham(m)$ at $m=1$ obtained from
the ten eigenvalues closest to zero on
50 gauge field configurations at $\beta=5.7$. The bin width 
is 0.005.
}
\label{fig:spec1}
\end{figure}

\begin{figure}
\epsfxsize=3.5in
\centerline{\epsffile{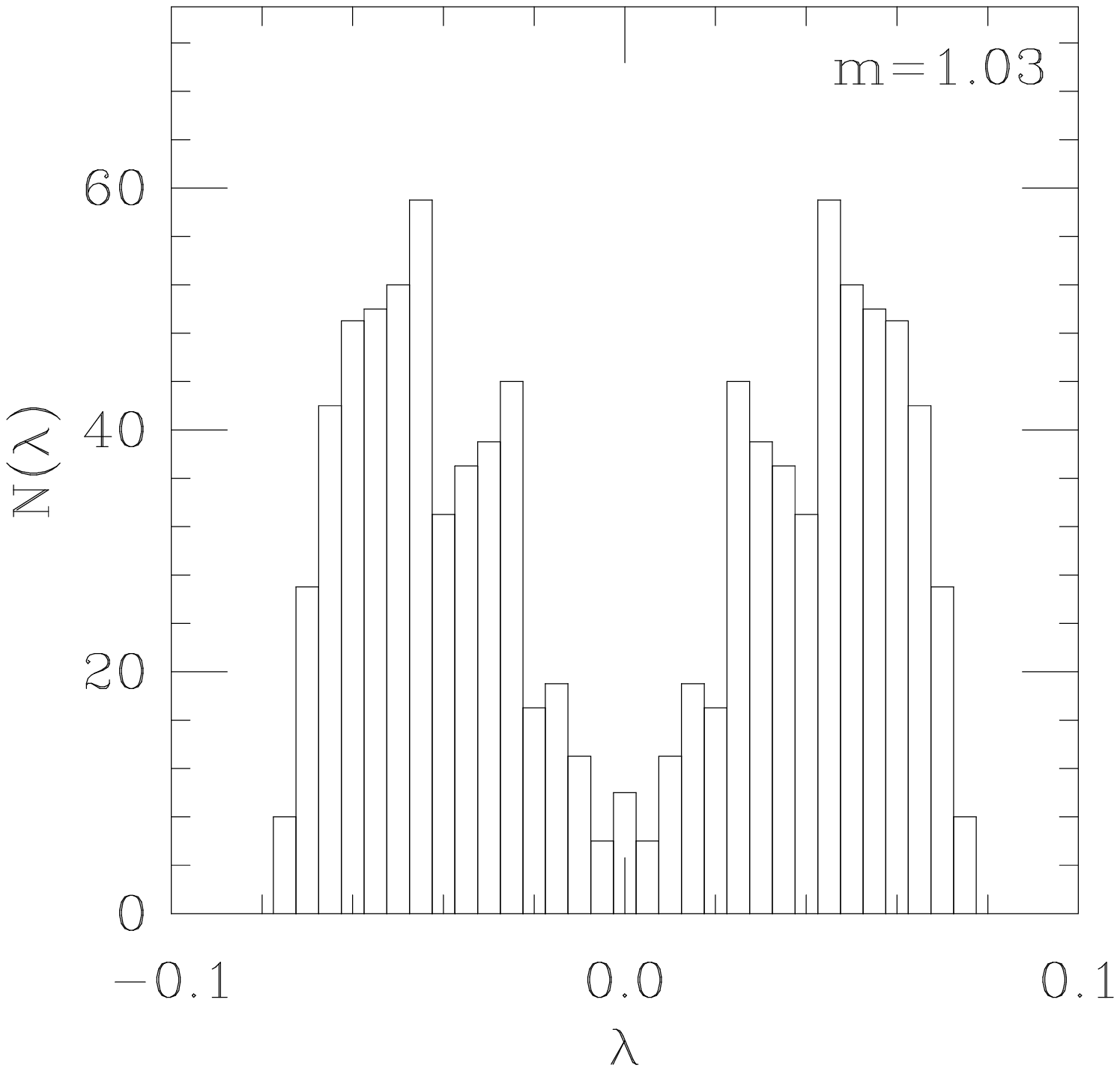}}
\caption{ \tenrm
Spectral distribution of $\ham(m)$ at $m=1.03$ obtained from
the ten eigenvalues closest to zero on
50 gauge field configurations at $\beta=5.7$. The bin width is
0.005.
}
\label{fig:spec2}
\end{figure}

We generated 50 independent SU(3) gauge field configurations 
on an $8^3\times 16$ lattice at $\beta=5.7$ using the standard
single plaquette Wilson action. We computed the ten eigenvalues
of $\ham(m)$ closest to zero as a function of $m$. For this,
we used the Ritz functional method developed in Ref.~\cite{ritz}
and improved for the case of multiple eigenvectors in Ref.~\cite{ks}.
The Ritz functional method finds the lowest eigenvalues of
$\ham^2(m)$, and these are in fact the eigenvalues of $\ham(m)$
closest to
zero. The ten eigenvalues of $\ham^2(m)$ were obtained to a
relative precision of $10^{-5}$ or to an absolute precision of
$10^{-9}$ depending on which condition is satisfied first in the
iterative procedure. Only ten eigenvectors were kept in the
iterative procedure and were reorthogonalized every ten iterations.
Eigenvalues and eigenvectors were obtained for masses between 
$m=1$ to $m=2$ with increments of $\Delta m=0.02$. 
In first order perturbation theory 
\begin{equation}
{d\lambda_n(m)\over dm} = - \phi^\dagger_n(m) \gamma_5 \phi_n(m) \ ,
\label{eq:perturb}
\end{equation}
with $\phi_n(m)$ being the eigenvector corresponding to
$\lambda_n(m)$. The mass increment was sufficiently small 
to allow for an accurate interpolation of eigenvalues 
between consecutive values of $m$. 
The ten eigenvalues closest to zero varied in
the range from $-0.1$ to $0.1$
as a function of $m$ over the entire ensemble.
Figure~\ref{fig:spec1} shows the spectral distribution 
$\rho(\lambda;m)$ at $m=1$. This distribution has a gap
and the physical quark is massive.
Figure~\ref{fig:spec2} shows the spectral distribution
$\rho(\lambda;m)$ at $m=1.03$. Here the spectrum does not
have a gap and the physical quark is massless.

\begin{figure}
\epsfxsize=3.5in
\centerline{\epsffile{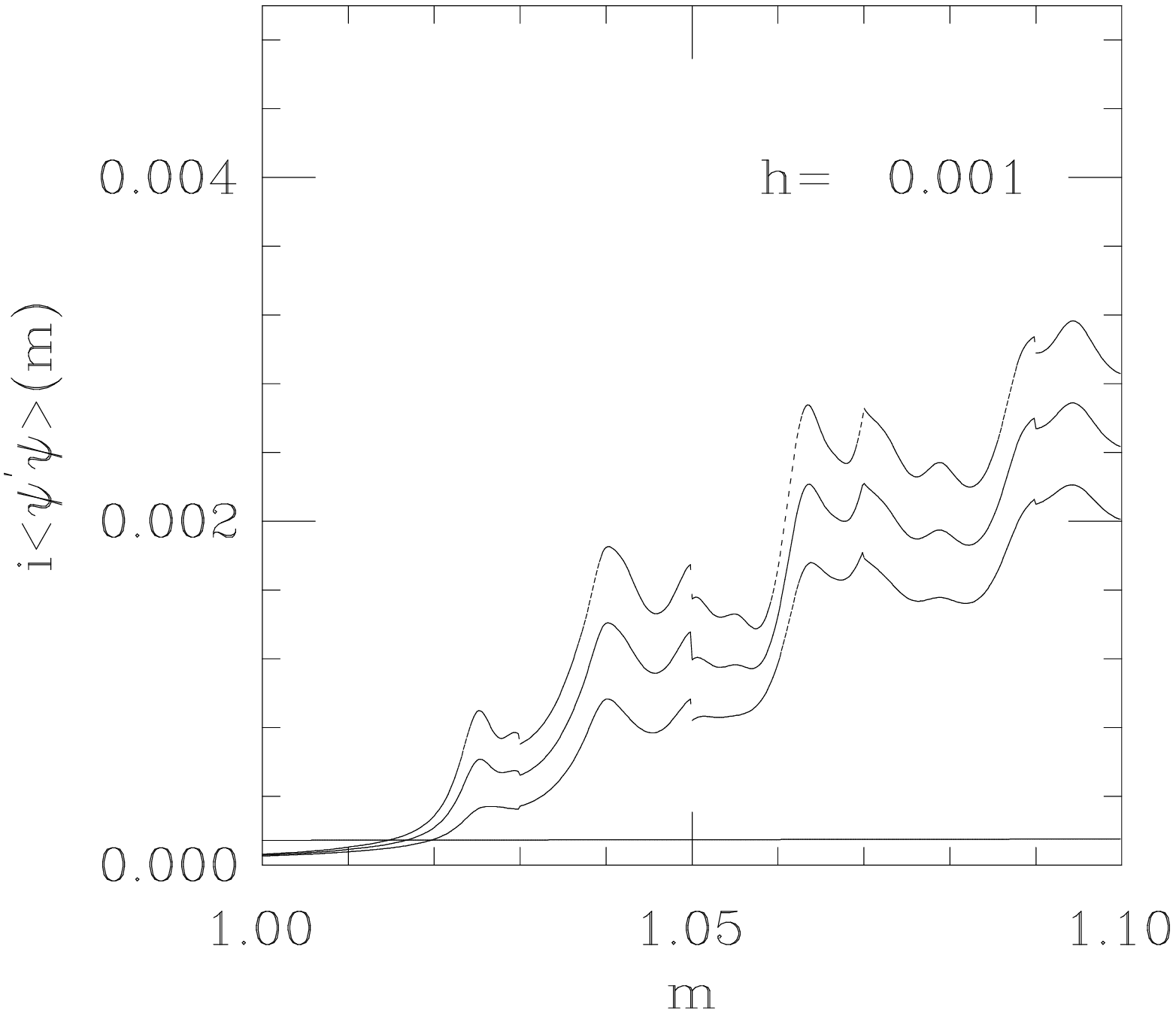}}
\caption{ \tenrm
The chiral condensate $i \langle\psi^\prime\psi\rangle(m)$
from $m=1.0$ to $m=1.1$ with $h=0.001$. 
The middle curve 
is the estimate of the chiral condensate, and the curves on either
side show the error of this estimate.
The solid line at the
bottom is the result for free fermions.
}
\label{fig:chiral1}
\end{figure}
\begin{figure}
\epsfxsize=3.5in
\centerline{\epsfbox{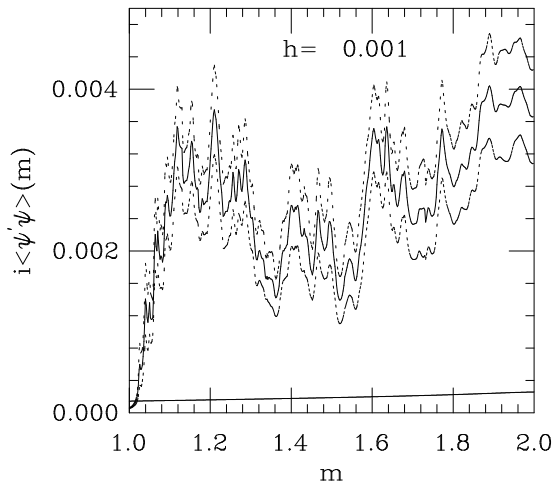}}
\caption{\tenrm
The chiral condensate $i\langle\psi^\prime\psi\rangle(m)$
from $m=1.0$ to $m=2.0$ with $h=0.001$. 
The middle curve
is the estimate of the chiral condensate, and the curves on either
side show the error of this estimate.
The solid line at the bottom is
the result for free fermions.
}
\label{fig:chiral2}
\end{figure}
Since $\rho(0;1.03)$ is non-zero, we expect a chiral condensate
at $m=1.03$. To investigate the chiral condensate as a function
of $m$, we use (\ref{eq:chiralU}) to measure
$i\langle\psi^\prime\psi\rangle_U$ per configuration and then
compute an ensemble average to obtain
$i\langle\psi^\prime\psi\rangle$. For small values of $h$ in
(\ref{eq:chiralU}), only the low lying eigenvalues will
contribute. Setting $h=0.001$ in (\ref{eq:chiralU}) and summing
over the ten lowest eigenvalues of $\ham(m)$, we obtain a value
for $i\langle\psi^\prime\psi\rangle(m)$ as a function of $m$. 
We also compute $i\langle\psi^\prime\psi\rangle(m)$ for free
fermions at $h=0.001$ by summing over the whole spectrum. 
We plot the result in the range $1\le m \le 1.1$
in Fig.~\ref{fig:chiral1}. The solid line at the bottom of the
figure is the result for free fermions. 
When \hbox{$m>1.02$},
the result from the gauge theory is significantly larger
than from free fermions, and is consistent with a
non-zero chiral condensate in this mass region.
Figure.~\ref{fig:chiral2} illustrates the behavior of the
condensate $i\langle\psi^\prime\psi\rangle(m)$ for the
region \hbox{$1.0<m<2.0$}. This result is consistent with 
a non-zero value for $\rho(0;m)$ for the entire range of 
masses between $m_1=1.02$ and $m_2=2.0$. Note that the critical
quark mass \hbox{$m_c=1.047$}, obtained by extrapolating the pion
mass to zero, is inside the region where the spectrum 
is gapless. 
\begin{figure}
\epsfxsize=3.5in
\centerline{\epsfbox{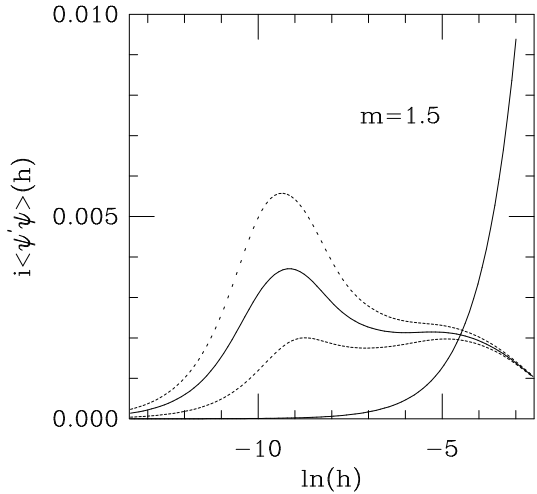}}
\caption{\tenrm
Chiral condensate $i\langle\psi^\prime\psi\rangle(h)$
at $m=1.5$ viewed as a function of $h$. The x-axis is logarithmic
to show a wide range in $h$. 
The middle curve with a peak
is the estimate of the chiral condensate, and the curves on either
side show the error of this estimate.
The exponentially varying solid line is            
the result for free fermions.
}
\label{fig:chiralh}
\end{figure}
To understand the dependence of the chiral condensate on the
parameter $h$, in Fig.~\ref{fig:chiralh} we plot
$i\langle\psi^\prime\psi\rangle$ versus $h$ at a typical
mass $m=1.5$. The picture is basically the same at other values
of $m$ greated than $m_1$. 
Clearly there is a wide region of $h$ where the
condensate is basically flat, in contrast to the free fermion
result which drops exponentially in the plot with a logarithmic scale.
As expected from (\ref{eq:chiralU}), 
the condensate goes to zero as $h$ goes to zero and
as $h$ goes to infinity.
The errors at the peak region are large
due to the large fluctuations in $\lambda_n^{-2}$ coming from
the low lying eigenvalues. If the spectral sum is restricted to the
six lowest eigenvalues, as opposed to the ten lowest, all the above
results are essentially unaltered.

\section {The mass range $\pmb{m_1 < m < m_c}$}

In the previous section we explored the consequences of the low
lying eigenvalues of $\ham(m)$. We showed that the spectrum is
gapless in a mass range bounded by $m_1=1.02$ and $m_2=2.0$, and
we provided evidence for a non-vanishing chiral condensate
in the gapless region. Our interpretation is that the physical quark
is massless in this whole region of $m$. If this is indeed the
case, we should find zero eigenvalues of $\ham(m)$ scattered
throughout this region due to the presence of topological
objects in the gauge field configurations. Equation
(\ref{eq:zero}) defines the zero eigenvalue equation for $\ham(m)$,
and (\ref{eq:size}) is a necessary condition for a zero
eigenvalue. Based on (\ref{eq:size}), it was argued in section 2
that larger topological objects will result in a zero eigenvalue
at smaller values of $m$. In this section we first look at the
sizes of the topological objects and their crossing points, and
we present evidence supporting the above statement. Then we
explore the region \hbox{$m_1<m<m_c$}, where $m_c=1.047$ is
the point at which the extrapolated pion mass vanishes, to
establish the following points:
\begin{itemize}
\item Eigenvectors associated with zero eigenvalues of $\ham(m)$
in the \hbox{$m_1<m<m_c$} region correspond to large localized objects
extending roughly over half the lattice in the long direction.
\item The pion propagator for a fixed gauge field background 
with level crossings, at values of $m$ close to a crossing point,
is well described by a spectral sum restricted to a few low lying
eigenvalues of $\ham(m)$. 
\item The ensemble average of the pion propagator obtained by
a restricted spectral sum at $m_1=1.02$ do not fit
an exponential decay. Instead they fit a power law.
\end{itemize}

To this end we write down the spectral representations of $\ham(m)$
and $\ham^{-1}(m)$. The eigenvalue equation for $\ham(m)$ is
\begin{equation}
\sum_{k^\prime bj}
\ham_{a i,b j}(k,k^\prime) \phi_{bj}^n(k^\prime) = \lambda_n 
\phi_{ai}^n(k) \ ,
\label{eq:eigen}
\end{equation}
and we have dropped the dependence on $m$ in the above equation.
Here $k$ and $k^\prime$ label the sites on the lattice, 
\hbox{$a$ and $b$} are four component spinor indices, \hbox{$i$
and $j$} are three component color indices, and $n$ labels the
eigenvalues. The eigenvectors are normalized to yield 
$\sum_{kai}[\phi^m_{a i}(k)]^* \phi^n_{a i}(k) = \delta_{nm}$. 
In terms of the eigenvalues and eigenvectors, the spectral
representations of $\ham$ and $\ham^{-1}$ are
\begin{equation}
\ham_{ai,bj}(k,k^\prime) = \sum_n \lambda_n
\phi^n_{a i}(k) [\phi^n_{b j}(k^\prime)]^*
\label{eq:hspec}
\end{equation}
\begin{equation}
\ham^{-1}_{ai,bj}(k,k^\prime) = \sum_n {1\over\lambda_n}\,
\phi^n_{a i}(k) [\phi^n_{b j}(k^\prime)]^* \ .
\label{eq:pspec}
\end{equation}
The point-like pion field is given by 
\begin{equation}
\pi(k) = \sum_{a i} \psi^\prime_{a i}(k) \psi_{a i}(k) \ ,
\label{eq:pion}
\end{equation}
and using the spectral representation of $\ham^{-1}$ in
(\ref{eq:pspec}), the pion propagator in a fixed gauge 
field background becomes
\begin{equation}
\langle \pi(k) \pi(k^\prime)\rangle_U = 
\sum_{nm} {1\over \lambda_n\lambda_m}
f_{nm}(k) f^*_{nm}(k^\prime) \ ,
\label{eq:pion_prop}
\end{equation}
where the gauge invariant functions $f_{nm}(k)$ are defined by
\begin{equation}
f_{nm}(k) = \sum_{a i} [\phi^n_{a i}(k)]^* \phi^m_{a i}(k) \ .
\label{eq:fnm}
\end{equation}

\begin{figure}
\epsfxsize=3.5in
\centerline{\epsffile{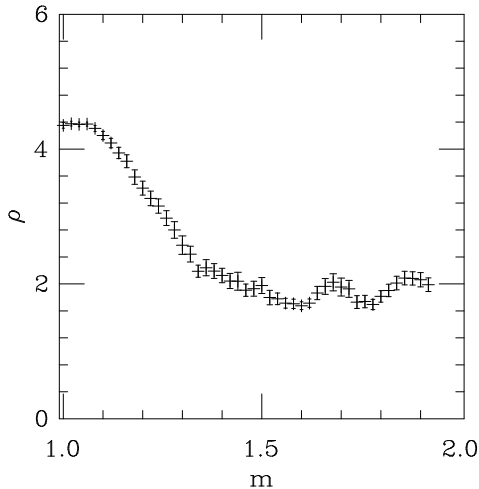}}
\caption{\tenrm
The localization size as a function of the
crossing point. The data point at $m$ includes all crossings
that occurred in the interval $m\pm 0.05$. 
}
\label{fig:rho_vs_mu}
\end{figure}
The eigenvalues of $H(m)$ will sweep out a
spectral flow as $m$ varies over its range, and typically some
levels will cross, thereby producing a set of zero modes at 
discrete values of the mass. Let us examine such a zero mode,
$\phi_{ai}^0(k)$, at some crossing point $m$. In what follows,
it is convenient to consider the zero momentum part of 
$f_{00}(k)$ defined by $f(t) = \sum_{\vec k} f_{00}(\vec k, t)$. 
We can associate a size with $f(t)$ using the pion correlator 
\begin{equation}
C_0(t) = \sum_{t^\prime} f(t^\prime) f(t+t^\prime) \ ,
\label{eq:zero_corr}
\end{equation}
which is a periodic function of $t$ with a peak at $t=0$. 
We define the localization size, $\rho$, through the second
moment of $C_0(t)$:
\begin{equation}
\rho^2 ={ \sum_{t=-7}^8 t^2 C_0(t) \over \sum_{t=-7}^8 C_0(t) } \ .
\label{eq:zero_size}
\end{equation}
Recall that the lattice has $16$ sites in the time direction, 
and in calculating
the localization size we order them to run from $-7$ to $8$.
Figure~\ref{fig:rho_vs_mu} shows the localization
size of the zero modes as a function of the location of the
crossing point. The trend in this figure is consistent with 
the picture that localized objects of a larger size cross
at smaller values of $m$. 

\begin{figure}
\epsfxsize=3.5in
\centerline{\epsffile{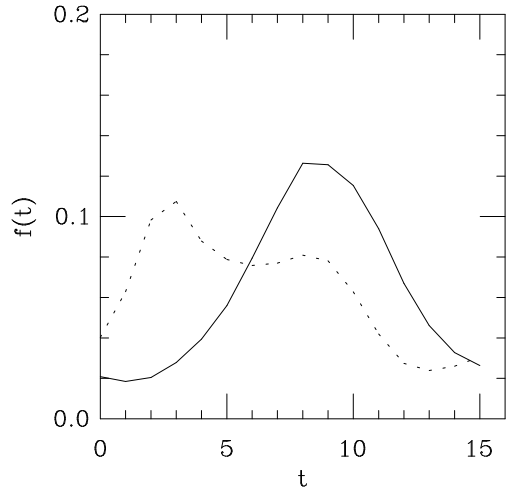}}
\caption{\tenrm
The function $f(t)$ for two near zero-modes within the mass
range \hbox{$m_1<m<m_c$}. The solid line corresponds to a
mode at $m=1.04$ for a certain configuration within the 
ensemble, while the dashed line corresponds to a mode at
$m=1.02$ for a different gauge field configuration.
}
\label{fig:zero}
\end{figure}
We now focus on crossings in the region between $m_1$ and $m_c$,
{\it i.e.} the mass range \hbox{$1.02 < m < 1.047$}. In our
ensemble of 50 configurations, we found seven which contained one
crossing each within this mass range, and one configuration 
with two crossings between $m_1$ and $m_c$.
For all nine zero modes,
$f(t)$ peaks at some value of $t$ in the range \hbox{$0 \le t \le
15$}, and in Fig.~\ref{fig:zero} we have plotted $f(t)$ for two
of them. In numerically solving for the eigenvalues of $H(m)$, we
generically find near zero-modes since we sweep over discrete values
of the mass parameter. The solid line in Fig.~\ref{fig:zero}
corresponds to a mode whose eigenvalue was closest to zero along
the computationally obtained discrete spectral flow. It occurred 
at a mass of $m=1.04$, and using (\ref{eq:perturb}) we find that
the associated zero crossing happens at $m_0=1.0375$, which is very
close indeed. As one can see, the mode peaks somewhere between
$t=8$ and $t=9$ and has a width of about seven or eight lattice
spacings. The dashed line is the eigenvector for another
configuration in the ensemble whose eigenvalue along the discrete
spectral flow is again closest to zero. It occurs at $m=1.02$, and
(\ref{eq:perturb}) implies that the associated zero crossing is
at $m_0=1.025$. This mode has a double peak structure,
and it stretches over ten lattice spacings. Of the nine zero
modes within the range $m_1<m<m_c$, seven of them had a single
peak in $f(t)$ while two exhibited double peaks, and all them
stretched over seven to ten lattice spacings. 

We also looked at $f_{00}(k)$ over the whole lattice and found
that the modes were localized in all four directions but were not
spherical. These modes are large objects having widths of seven to
ten lattice spacings in the long direction and widths of six or
so in the short directions. The finite size of the lattice plays
a significant role in the shape of the mode. Furthermore, the
associated lattice gauge field backgrounds of these eigenvectors
are far from the continuum limit, and as such are not smooth
configurations. This also affects the shape of these modes.

\begin{figure}
\epsfxsize=3.5in
\centerline{\epsffile{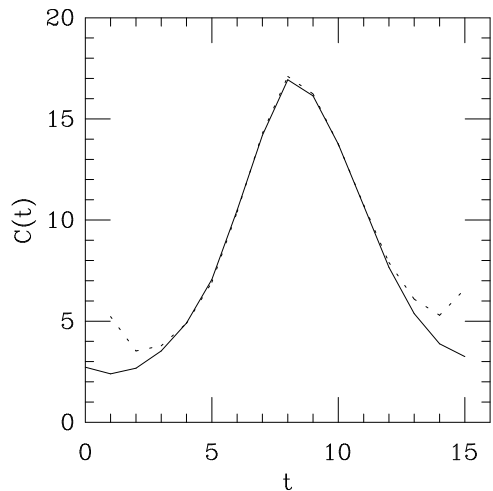}}
\caption{\tenrm
Comparison between the full pion propagator (dashed) and the pion
propagator obtained from the restricted spectral sum (solid), both
of which are in the fixed gauge field background corresponding to
the solid line of Fig.~7. The mass is $m=1.029$, which
is near the crossing value $m_0=1.0375$.
}
\label{fig:pion1}
\end{figure}
\begin{figure}
\epsfxsize=3.5in
\centerline{\epsffile{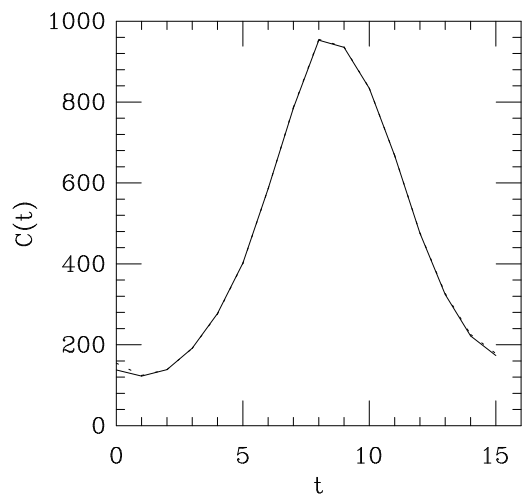}}
\caption{\tenrm
Comparison between the full pion propagator (dashed) and the pion
propagator obtained from the restricted spectral sum (solid) using
the same gauge field background as in Fig.~8. This time the mass
is $m=1.036$, which is even closer to the crossing value
$m_0=1.0375$, and the two curves lie almost on top of one another.
}
\label{fig:pion2}
\end{figure}
Near values of $m$ where $\ham(m)$ has a zero eigenvalue, we expect
the spectral decomposition of $\ham^{-1}(m)$, as given by (\ref{eq:pspec}),
to be dominated by the low lying 
modes. We test this on the
configuration corresponding to the solid line in Fig.~\ref{fig:zero}
at two mass points $m=1.029$ and $m=1.036$, the latter being
close to the crossing point $m_0=1.0375$. We shall restrict the
spectral sum in (\ref{eq:pion_prop}) to run only over the ten
eigenvalues closest to zero, and we compute the quantity $C(t) =
\sum_{\vec k} \langle \pi(0) \pi(\vec k, t) \rangle_U$ using both
the restricted spectral sum and the full propagator obtained by
inverting $\ham(m)$ with standard techniques.
Fig.~\ref{fig:pion1} shows $C(t)$ at $m=1.029$ and
Fig.~\ref{fig:pion2} illustrates a comparison at $m=1.036$. 
At $m=1.029$  both methods agree for \hbox{$3\le t \le 12$}, while
at $m=1.036$ they agree at all values of $t$, which is an expected
result since the latter mass is closer to the crossing point.
Furthermore, it should come as no surprise that the pion propagator
itself, $C(t)$, looks very much like $f(t)$ for the corresponding
zero mode, in this case the solid line of Fig.~\ref{fig:zero}.

\begin{figure}
\epsfxsize=3.5in
\centerline{\epsffile{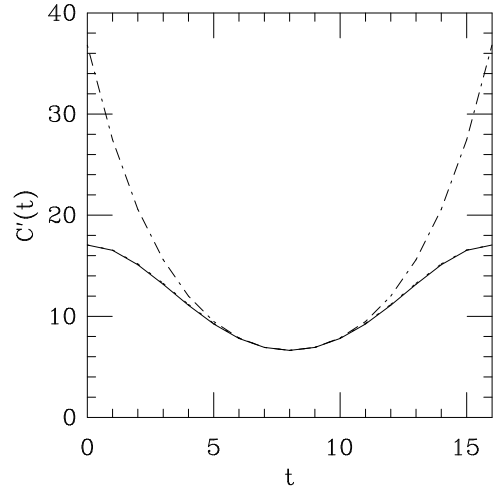}}
\caption{\tenrm 
The solid line is the
pion propagator of Fig.~8 
averaged over
all possible translations of the gauge field. The dotted
line that falls on top of the solid line is a fit to a
power law, and the dashed line shows an exponential fall 
off in comparison.
}
\label{fig:pion_av}
\end{figure}

The pion propagators of Figs.~\ref{fig:pion1} and \ref{fig:pion2}
are far away from an expected exponential decay starting at $t=0$,
and instead they have a dominant peak around $t=8$. Such
configurations are the reason why one cannot measure the pion
mass at values of $m$ in the region between $m_1=1.02$ and
$m_c=1.047$, since they cause large statistical fluctuations.
This has been known for quite some time and was first observed
in Ref.~\cite{original}. As previously mentioned, the peak is due 
to a zero of $\ham(m)$ close to values of $m$ where the pion
propagator is being measured. Since we have presented evidence
that the pion propagator can be well approximated by a restricted
spectral sum, one can reduce the statistical fluctuations by
computing the pion propagator as an average over all possible
translations of the background gauge field. This can easily be
done in the restricted spectral sum, but it would be prohibitive
if one were using conventional methods to compute the pion
propagator, since one would have to do a volume worth of inversions of
the Wilson-Dirac operator. It should be mentioned that a
restricted spectral sum has been used to study the spectroscopy
in the region where $\ham(m)$ has a gap~\cite{Negele}. The
average pion propagator over all possible translations of a fixed
gauge field background is 
\begin{equation}
C_{\rm av}(k) = {1\over V} \sum_{k^\prime} 
\langle \pi(k^\prime)\pi(k+k^\prime)\rangle \ ,
\label{eq:pion_av}
\end{equation}
and we find it useful to study $C^\prime (t) = \sum_{\vec k}
C_{av}(\vec k,t)$. In Fig.~\ref{fig:pion_av} we plot $C^\prime
(t)$ at $m=1.029$ for the same configuration as in
Fig.~\ref{fig:pion1}. This function has the expected behavior for
a pion correlator in the sense that it falls off starting at
$t=0$, but it does not fit an exponential decay. Instead, it fits
a power law of the form obtained using the zero mode solution in
the presence of a classical instanton. The spherically symmetric
zero mode associated with a classical instanton take the 
form~\cite{Hooft}
\begin{equation}
\psi^\dagger(x) \psi(x) \sim {1\over (r^2 + \rho^2)^3} \ ,
\label{eq:zero_mode}
\end{equation}
which gives a contribution to the pion correlator of
\begin{equation}
C_{\rm cl}(x) = \int {d^4y\over (2\pi)^4} {1\over (y^2 + \rho^2)^3}
{1\over ((x+y)^2 + \rho^2)^3} \ .
\label{eq:pion_zero_mode}
\end{equation}
\begin{figure}
\epsfxsize=3.5in
\centerline{\epsffile{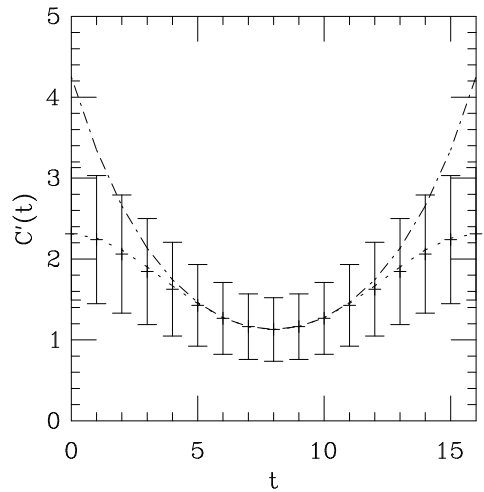}}
\caption{\tenrm
The data points with error bars show the 
 pion correlator $C^\prime(t)$ averaged over the
ensemble of 50 configurations.
The dotted
line is a power law decay and
the dashed line shows an exponential fall off
in comparison.
}
\label{fig:pion_ave}
\end{figure}

The above integral can be evaluated and the resulting functional
form can be fitted to $C^\prime(t)$. The fit corresponding to 
$\rho=5.1375$ is shown as the dotted line in Fig.~\ref{fig:pion_av}, 
which falls right on top of $C^\prime(t)$ represented by the
solid line. 
In contrast to this fit, the dashed line shows the
behavior of  $C(e^{-\alpha t} + e^{-\alpha (16-t)})$ with
$\alpha=0.3$. Note that $C^\prime(t)$ is below the exponential
decay near $t=0$ implying that the decay is slower at $t=0$ when
compared to the decay at $t=8$. This is atypical behavior for the
correlator since excited states would result in a faster decay
at $t=0$. This is another indication that the behavior is
inconsistent with an exponential decay.
The power law behavior is consistent with a 
non-vanishing chiral condensate in the region \hbox{$1.02 < m <
1.047$}, as found in the previous section. Furthermore,
Fig.~{\ref{fig:zero} shows that the zero eigenvalues in the above
region of $m$ are caused by large topological objects extending
over several lattice spacings. Therefore, zero eigenvalues are
caused by physical objects in the background gauge field, which
results in a non-vanishing chiral condensate and a power law
behavior in the pion mass. The proposal of Ref.~\cite{fermi}
shifts the position of these zeros to some value of $m$ greater
than $m_c=1.047$, thereby removing all low lying eigenvalues 
in the region \hbox{$m < m_c$}. This should also make the chiral
condensate vanish for \hbox{$m < m_c$} and remove the power law
contribution to the pion propagator in this region. This
indicates that the proposal in Ref.~\cite{fermi} is a non-local
change to the Wilson-Dirac operator motivated by the need for a
single value of $m$ where the pion mass vanishes. The proposal
of Ref.~\cite{fermi} does nothing to the zero crossings that
occur at $m>m_c$, and therefore one will still have a region
where $\ham(m)$ is gapless. In this region there will be a
non-vanishing chiral condensate and the pion will be massless. In
effect, the modification of the eigenvalues proposed
in Ref.~\cite{fermi} slightly changes the region where the
spectrum of $\ham(m)$ is gapless and we see no compelling
reason to make this non-local modification to the Wilson-Dirac
operator.

We now proceed to compute $C^\prime (t)$ as an average over the
whole ensemble at $m_1=1.02$. At this mass value there is
evidence for a non-zero density of eigenvalues at zero, as 
presented is section 3. Figure~\ref{fig:pion_ave} shows the
result of averaging over the whole ensemble. The errors, although
large, still enable us reasonably estimate of the pion
correlator at this value of $m$. The dotted line shows the same
power law function as the one used in Fig.~\ref{fig:pion_av} with
$\rho=5.675$, whereas the dashed line shows an exponential decay
with $\alpha=0.25$. Again we see the same behavior as in 
Fig.~\ref{fig:pion_av}, namely the data is below the exponential decay
at $t=0$. The pion propagator at different times is correlated and
it is interesting to note that the power law fit goes through all
the central values. 
The propagator fits the power law better and is
not consistent with an exponential decay. 
We have therefore shown
that it is not appropriate to alter the low lying eigenvalues in
the region of m between $m_1=1.02$ and $m_c=1.047$. We have also
established that (\ref{eq:pion_av}) is a powerful method for
estimating the pion propagator in this region using the
restricted spectral sum to compute the right hand side of this
equation. 

\section{Conclusions}

We have studied the spectrum of $\ham(m)=\gamma_5\wilson(-m)$,
with $\wilson(m)$ being the Wilson-Dirac operator on the lattice
with a bare mass $m$. We generated 50 pure $SU(3)$ gauge configurations
using the single plaquette Wilson action at $\beta=5.7$
on an $8^3\times 16$ lattice. We found that the spectrum is
consistent with it being gapless in the region between $m_1=1.02$
and $m_2=2.0$. Using a parity breaking operator as proposed in
Ref.~\cite{pfb}, we estimated the chiral condensate and found it
to be non-zero in this whole region. Therefore, it is natural to
say that the physical quark is massless in this region.
This is in contrast to declaring a single value, $m_c =1.047$, where the
pion mass obtained by extrapolation from the region $m < 1$ 
vanishes~\cite{gf11},
as the unique point where the physical quark is massless. 
We computed the pion
correlator using a restricted spectral sum and by averaging over
all possible translations of the background gauge field. We found
that the pion correlator is consistent with a power law decay at
$m_1=1.02$, with the power law coming from the low lying
eigenvalues of $\ham(m)$. This further confirms that the physical
quark is massless in this whole region of $m$, as opposed to a
single point. We also showed that the eigenvectors corresponding
to zero eigenvalues of $\ham(m)$ in the region between $m_1=1.02$
and $m_c=1.047$ are localized and roughly span half the lattice
in the long direction (about 1.36~fm). As such, the low lying
eigenvalues in this region are associated with physical objects
in the gauge field background. 

The region in $m$ where
the physical quark is massless shrinks to the single point $m=0$
in the continuum limit. But the region has a finite width if
$\beta$ is finite. The width will change if we improve the
Wilson-Dirac operator. We studied the spectral flow for a few
gauge field configurations using the
Sheikholeslami-Wohlert~\cite{clover} improvement scheme. We found
that the gap closes at a smaller value of $m$, but preliminary
indications
are that it remains closed all the way up to $m=2$, as in the
Wilson-Dirac case. For the Wilson-Dirac operator, the width of
the gapless region is not expected to change as we go to infinite
lattice volume at fixed $\beta$. But we expect more low lying
eigenvalues as we go to larger lattice volume since the chiral
condensate arises from spontaneous symmetry breaking, and the
effect should become stronger in larger lattice volume. For the
same reason, one should see more zero eigenvalues of $\ham(m)$ in
a fixed region of $m$. An investigation of ten configurations on
a $16^3\times 32$ lattice at $\beta=5.7$ showed that there were
seven zero eigenvalues in the 
region \hbox{$1.02 < m < 1.047$}. 
This is a greater percentage than the nine zero eigenvalues found
in 50 configurations on the $8^3
\times 16$ lattice at the same coupling. The eigenvectors
associated with these modes were found to be localized and large
just like on the $8^3\times 16$ lattice. There will be two
competing effects as one goes to larger volume. As remarked, one
will find more zero modes in the region $1.02 < m < 1.047$. But
the effect of a single zero mode in larger volume will not be
felt in as wide a region around the crossing point as in a smaller
volume. This is because the number of modes that
contribute to the pion propagator increase with volume
and each single mode therefor has less weight.

Since for $\beta=5.7$ the gap does not reopen, the natural
conclusion is that we are far from the continuum limit. We
expect the gap to open again before $m=2$ if we go to a larger
$\beta$. Tracing out the gapless region in the $(\beta, m)$ plane
is necessary to probe massless quarks as one goes to the
continuum limit. It is also important for a proper study of
massless QCD using the overlap formalism~\cite{over} or the 
domain wall formalism~\cite{domain}. For a proper study one
has to work at a value of $\beta$ so that the gapless region ends before
$m=2$.To reproduce the proper chiral behavior, the value of the mass that
goes into the overlap Hamiltonian in (\ref{eq:many_body}), and
the domain wall mass, should be kept greater than the upper
bound of the gap $m_2(\beta)$. 
The
gap also has to open before $m=2$ so that one can
use level crossings for a proper measurement of the
topological charge and 
topological susceptibility. 

\ack{
This research was supported by DOE contracts 
DE-FG05-85ER250000, DE-FG05-96ER40979 and DE-FG03-96ER40956.
R.L.S would like to thank Steve Sharpe and Larry Yaffe and
R.G.E, U.M.H and R.N would like to thank Tony Kennedy,
Tim Klassen and Stefan Sint for useful discussions.
Computations were performed on the CM-2,
workstation cluster and the new QCDSP supercomputer at SCRI.}


\begin{thebibliography}{9}

\bibitem{pion} 
S. Gottlieb, {\em Nucl. Phys. \/} {\bf B53} (Proc. Suppl.) (1997) 155.
\bibitem{bc}
T. Banks and A. Casher, {\em Nucl. Phys. \/} {\bf B169} (1980) 103.
\bibitem{index}
M. Atiyah and I. Singer, {\em Ann. Math. \/} {\bf 87} (1968) 484.
\bibitem{over} R. Narayanan and H. Neuberger, 
{\em Phys. Rev. Lett.  \/}{\bf 71} (1993) 3251;
{\em Nucl. Phys.  \/}{\bf B443} (1995) 305. 
\bibitem{su2_top} P. Vranas and R. Narayanan, hep-lat/9702005, to
appear in {\em Nucl. Phys. B. \/}
\bibitem{pfb} K.M. Bitar, U.M. Heller and R. Narayanan, hep-th/9710052.
\bibitem{fermi} W. Bardeen, A. Duncan, E. Eichten, G. Hockney
and H. Thacker, hep-lat/9705008, hep-lat/9710084.
\bibitem{gf11} F. Butler, H. Chen, J. Sexton, A. Vaccarino and
D. Weingarten, {\em Nucl. Phys.  \/}{\bf B430} (1994) 179.
\bibitem{ritz} B. Bunk, K. Jansen, M. L\"uscher and H. Simma,
DESY-Report (September 1994).
\bibitem{ks} T. Kalkreuter and H. Simma,{\em  Comput. Phys. Commun. \/}
 {\bf 93} (1996) 33.
\bibitem{original} K.-H. Mutter, Ph. De Forcrand, K. Schilling and
R. Sommer, in Brookhaven 1986, 
{\em Proceedings, Lattice Gauge Theory, '86, \/}
pg. 257.
\bibitem{Negele} M.-C. Chu, J.M. Grandy, S. Huang and J.W. Negele,
{\em Phys. Rev. Lett. \/}{\bf 70} (1993) 225; 
{\em Phys. Rev.  \/}{\bf D48} (1993) 3340;
{\em Phys. Rev.  \/}{\bf D49} (1994) 6039;
T.L. Ivanenko and J.W. Negele, hep-lat/9709130.
\bibitem{Hooft} G. 't Hooft, {\em Phys. Rev.  \/}{\bf D14} (1976) 3432.
\bibitem{clover} B.-Sheikholeslami and R. Wohlert, 
{\em Nucl. Phys.  \/}{\bf B259} (1985) 572.
\bibitem{domain} D.B. Kaplan, {\em Phys. Lett.  \/}{\bf B288} (1992) 342,
T. Blum and A. Soni, hep-lat/9611030; hep-lat/9706023;
hep-lat/9710051.
\end{thebibliography}
\end{document}